\documentclass{elsart}
\usepackage{graphicx, amssymb}
\newcommand{\D}{\displaystyle}

\makeatletter
 \newcommand{\figcaption}{\def\@captype{figure}\caption}
 \newcommand{\tabcaption}{\def\@captype{table}\caption}
\makeatother

\begin{document}

\begin{frontmatter}

\title {\bf Proton and neutron skins of light nuclei \\
within the Relativistic Mean Field theory}

\author[rcnp,beijing]{L.S. Geng},
\author[rcnp,riken]{H. Toki},
\author[riken]{A. Ozawa},
\author[beijing]{J. Meng}



\address[rcnp]{Research Center for Nuclear Physics (RCNP), Osaka
University,\\
Ibaraki, Osaka 567-0047, Japan}

\address[riken]{The Institute of Physical and Chemical Research
(RIKEN),\\
Wako, Saitama 351-0198, Japan}

\address[beijing]{School of Physics, Peking University, Beijing 100871, P.R.
China }

\begin{abstract}
The Relativistic Mean Field (RMF) theory is applied to the
analysis of ground-state properties of Ne, Na, Cl and Ar isotopes.
In particular, we study the recently established proton skin in Ar
isotopes and neutron skin in Na isotopes as a function of the
difference between the proton and the neutron separation energy.
We use the TMA effective interaction in the RMF Lagrangian, and
describe pairing correlation by the density-independent
delta-function interaction. We calculate single neutron and proton
separation energies, quadrupole deformations, nuclear matter radii
and differences between proton radii and neutron radii, and
compare these results with the recent experimental data.
\end{abstract}

\end{frontmatter}

\section {Introduction}
One of the most fundamental quantity of the nucleus is its size,
nuclear radius, which provides important information on the effect
of the nuclear potential, shell effect and ground-state
deformation. The presence of neutron skin in stable nuclei has
been discussed since the late 1960s \cite{myers.69}. No evidence
of appreciable neutron skin in stable nuclei has been observed. A
clear evidence has not been obtained even for neutron-rich nuclei
until 1990s. During the last decade, developments in accelerator
technology and detection techniques \cite{tanihata.95} have led to
an extended study of exotic nuclei near the limits of the $\beta$-
stability, especially those light nuclei with $4\leq Z\leq 12$. In
particular, measurements of interaction cross sections
$(\sigma_I)$ for light nuclei have been extensively studied at
many radioactive nuclear beam facilities and provided important
data on nuclear radii
\cite{suzuki.95,suzuki.98,chulk.96,ozawa.01,ozawa.02}.

In very neutron-rich nuclei, the weak binding of the outermost
neutrons causes the formation of the neutron skin on the surface
of a nucleus, and the formation of one- and two-neutron halo
structures. Measurements of Na isotopes \cite{suzuki.95,suzuki.98}
have revealed a monotonic increase in the neutron skin thickness
with increasing neutron number. The established two-neutron halo
nuclei are $^6$He, $^{11}$Li and $^{14}$Be, and the one-neutron
halo nuclei are $^{11}$Be and $^{19}$C. Recent data
\cite{ozawa.01} present evidence for a one-neutron halo in
$^{22}$N, $^{23}$O and $^{23}$F. On the proton-rich side evidence
has been reported for the existence of a proton skin in $^{20}$Mg
\cite{chulk.96}. More recently, Ozawa et al. \cite{ozawa.02} has
confirmed for the first time a monotonic increase of proton skin.
These observations indicate that neutron as well as proton skin
are quite common phenomena in unstable nuclei far from the
stability line. Data on ground-state deformations are also very
important for the study of shell effect in exotic nuclei. In
particular, they reflect major modifications in the shell
structures, the disappearance of traditional and the occurrence of
new magic numbers. The new magic number $N=16$ has been reported
in the light neutron drip line region \cite{ozawa.00}. Different
ground-state deformations of proton and neutron density
distributions are expected in some nuclei with extreme isospin
projection quantum number, $T_z$.

The Relativistic Mean Field (RMF) theory has been successfully
applied to study global properties of nuclei from the proton drip
line to the neutron drip line in the entire mass region
\cite{sugahara.94,ring.96}. The parameter sets, TM1 and TM2, were
invented to describe ground-state properties of heavy nuclei
($A>40$; TM1) and light nuclei ($A<40$; TM2) \cite{sugahara.94}.
The parameter set, TM1, has a favorable property of being able to
reproduce the essential feature of the equation of state and the
vector and the scalar self-energies of the relativistic
Bruckner-Hartree-Fock theory for nuclear matter \cite{brock.90}.
This parameter set was, however, not able to provide good
quantitative results for light nuclei, and the introduction of the
other parameter set, TM2, was forced to be made. Considering the
RMF Lagrangian with the model parameters as completely a
phenomenological model for description of nuclear ground states
within the RMF approximation, Sugahara introduced a parameter set
smoothly varying with mass and named it as TMA \cite{sugahara.95},
which interpolates the TM1 and the TM2 parameter sets with very
week mass dependence. We may interpret this mass dependence as a
mean to effectively express the quantum fluctuations beyond the
mean field level and/or the softness of the nuclear ground states
in deformation, pairing and alpha clustering in light nuclei. On
the other side, the parameter set, NL3 \cite{ring.97}, is another
very successful parameter set for the description of nuclear
properties in the entire mass region, which originates from the
original non-linear parameter set, NL1 \cite{rein.86}. The NL3
parameter set has removed the unwanted features of NL1 of not
being able to obtain stable states for $^{12}$C and $^{16}$O due
to too large attractive contribution of the non-linear terms,
while keeping the good properties of the NL1 parameter set for
heavy nuclei. It is, therefore, very interesting to compare the
theoretical results of the RMF model with these parameter sets,
TMA, TM2 and NL3 with the newly obtained systematic data on the
light mass region.

In the present work, we apply the recently developed deformed
RMF+BCS method with a density-independent delta-function
interaction in the pairing channel \cite{geng.03} to the analysis
of ground-state properties of Ne, Na, Cl and Ar isotopes. As for
the mean field part, we use the TMA, TM2 and the NL3 parameter
sets and compare, in particular, the proton and the neutron skins
in Ar and Na isotopes. We compare also with the HF+BCS model with
the Skyme parameter \cite{gori.01}. We note that there is a work
by Lalazissis et al. \cite{lala.01} studying deformed light
nuclei, Be, B, C, N, F, Ne and Na isotopes with the Relativistic
Hartree-Bogoliubov (RHB) method.

In Sect. 2, we shall present the RMF theory with deformation and
pairing correlation.  We provide the numerical results in Sect. 3
for Ne, Na, Cl and Ar isotopes.  Sect. 4 is devoted to the summery
of the present study.

\section{Relativistic mean field theory with deformation and
pairing}

Our RMF calculations have been carried out using the model
Lagrangian density with nonlinear terms for both $\sigma$ and
$\omega$ mesons as described in detail in Ref. \cite{geng.03},
which is given by
\begin{equation}
\begin{array}{lll}
\mathcal{L} &=& \bar \psi (i\gamma^\mu\partial_\mu -M) \psi +
\,\frac{\D 1}{\D 2}\partial_\mu\sigma\partial^\mu\sigma-\frac{\D
1}{\D 2}m_{\sigma}^{2} \sigma^{2}- \frac{\D 1}{ \D
3}g_{2}\sigma^{3}-\frac{\D 1}{\D
4}g_{3}\sigma^{4}-g_{\sigma}\bar\psi
\sigma \psi\\
&&-\frac{\D 1}{\D 4}\Omega_{\mu\nu}\Omega^{\mu\nu}+\frac{\D 1}{\D
2}m_\omega^2\omega_\mu\omega^\mu +\frac{\D 1}{\D
4}g_4(\omega_\mu\omega^\mu)^2-g_{\omega}\bar\psi
\gamma^\mu \psi\omega_\mu\\
 && -\frac{\D 1}{\D 4}{R^a}_{\mu\nu}{R^a}^{\mu\nu} +
 \frac{\D 1}{\D 2}m_{\rho}^{2}
 \rho^a_{\mu}\rho^{a\mu}
     -g_{\rho}\bar\psi\gamma_\mu\tau^a \psi\rho^{\mu a} \\
      && -\frac{\D 1}{\D 4}F_{\mu\nu}F^{\mu\nu} -e \bar\psi
      \gamma_\mu\frac{\D 1-\tau_3}{\D 2}A^\mu
      \psi,\\
\end{array}
\end{equation}
where the field tensors of the vector mesons and of the
electromagnetic field take the following forms:
\begin{equation}
\left\{
\begin {array}{rl}
\Omega_{\mu\nu} =&
\partial_{\mu}\omega_{\nu}-\partial_{\nu}\omega_{\mu}\\
 R^a_{\mu\nu} =& \partial_{\mu}
                  \rho^a_{\nu}
                  -\partial_{\nu}
                  \rho^a_{\mu}-2g_\rho\epsilon^{abc}\rho^b_\mu\rho^c_\nu\\
 F_{\mu\nu} =& \partial_{\mu}A_{\nu}-\partial_{\nu}
A_{\mu}
\end{array}\right.
\end{equation}
and other symbols have their usual meanings. Based on the
single-particle spectrum calculated by the RMF method described
above, we perform a state-dependent BCS calculation
\cite{lane.64,ring.80}. The gap equation has a standard form for
all the single particle states. i.e.
\begin{equation}\label{eq:bcs}
\Delta_k=-\frac{\D 1}{\D 2}\sum_{k'>0}\frac{\D \bar{V}_{kk'}
\Delta_{k'}}{\D\sqrt{(\varepsilon_{k'}-\lambda)^2+\Delta_{k'}^2}},
\end{equation}
where $\varepsilon_{k'}$ is the single particle energy and
$\lambda$ is the Fermi energy. The particle number condition is
given by $2\sum\limits_{k>0} v_k^2=N$. In the present work we use
for the pairing interaction a delta-function interaction, i.e.
\begin{equation}
V=-V_0\delta({\bf r_1}-{\bf r_2}),
\end{equation}
with the same strength $V_0$ for both protons and neutrons. The
pairing matrix element for the delta-function interaction is given
by
\begin{equation}\label{eq:me}
\begin{array} {lll}\bar{V}_{ij}&=&\langle i\bar{i}|V|j\bar{j}\rangle-\langle i\bar{i}|V|\bar{j}j\rangle=-V_0\int d^3r\,
\left[\psi^\dagger_i\psi^\dagger_{\bar{i}}\psi_j\psi_{\bar{j}}-\psi^\dagger_i\psi^\dagger_{\bar{i}}\psi_{\bar{j}}\psi_j\right]\\
\end{array}
\end{equation}
with nucleon wave function in the form
\begin{equation}\label{eq:spinor}
\psi_i({\bf r},t)=\left(\begin{array}{c}f_i({\bf r})\\ig_i({\bf
r})\end{array}\right) =\frac{\D
1}{\D\sqrt{2\pi}}\left(\begin{array}{l}f^+_i(z,r_\bot)e^{i(\Omega_i-1/2)\varphi}\\
f^-_i(z,r_\bot)e^{i(\Omega_i+1/2)\varphi}\\ig^+_i(z,r_\bot)e^{i(\Omega_i-1/2)\varphi}\\
ig^-_i(z,r_\bot)e^{i(\Omega_i+1/2)\varphi}\end{array}\right)\chi_{t_i}(t).
\end{equation}
A detailed description of the deformed RMF+BCS method can be found
in Ref. \cite{geng.03}.

In the present study, nuclei with both even and odd number of
protons (neutrons) need to be calculated. We adopt a simple
blocking method without breaking the time reversal symmetry. The
ground state of an odd system is described by the wave function,
\begin{equation}
\alpha^\dagger_{k_1}|BCS>=\alpha^\dagger_{k_1}\prod_{k\ne
k_1}(u_k+v_k\alpha^\dagger_k\alpha^\dagger_{\bar{k}})|vac>.
\end{equation}
Here, $|vac>$ denotes the vacuum state.  The unpaired particle
sits in the level $k_1$ and blocks this level. The Pauli principle
prevents this level from participating in the scattering process
of nucleons caused by the pairing correlations. As described in
Ref. \cite{ring.80}, in the calculation of the gap, one level is
"blocked":
\begin{equation}
\Delta_k=-\frac{\D 1}{\D 2}\sum_{k'\ne k_1>0}\frac{\D
\bar{V}_{kk'}
\Delta_{k'}}{\D\sqrt{(\varepsilon_{k'}-\lambda)^2+\Delta_{k'}^2}}.
\end{equation}
The level $k_1$ has to be excluded from the sum because it can not
contribute to the pairing energy. The corresponding chemical
potential is determined by
\begin{equation}
N=1+2\sum_{k\ne k_1>0}v^2_k.
\end{equation}
The blocking procedure is performed at each step of the
self-consistent iteration.

\section{Ground state properties of light nuclei}

In the present work we take the mass-dependent parameter set TMA
\cite{sugahara.94} for the RMF Lagrangian. Calculations with
parameter set TM2 \cite{sugahara.94} and NL3 \cite{ring.97} are
also performed for comparison. In the pairing channel, pairing
strength $V_0=343.7$ Mev fm$^3$ \cite{geng.03} is taken for both
protons and neutrons in the density-independent delta-function
interaction. For each nucleus, first the constrained quadrupole
calculation \cite{flocard.73} is done to obtain all possible
ground-state configurations; second we perform the non-constrained
calculation using the quadrupole deformation parameter of the
deepest minimum of the energy curve of each nucleus as the
deformation parameter for our Harmonic Oscillator basis. In the
case of several similar minima from the constrained calculation,
we repeat the above procedure to botain the configuration with the
lowest energy as our final result. The calculations for the
present analysis have been carried out by an expansion in 14
oscillator shells for both fermion fields and boson fields.
Convergence has been checked with more shells. Following Ref.
\cite{gambhir.90}, we fix $\hbar\omega_0=41A^{-1/3}$ for fermions.
In what follows we discuss the details of our calculations and the
numerical results.

\subsection{Single neutron and proton separation energy}
{\noindent Single neutron separation energy:}
\begin{equation}
S_n(Z,N)=B(Z,N)-B(Z,N-1)
\end{equation}
 and single proton
separation energy:
\begin{equation}
\label{eq:sp}
 S_p(Z,N)=B(Z,N)-B(Z-1,N)
\end{equation}
are very important and sensitive quantities of nuclei, where
$B(Z,N)$ are the binding energies of nuclei with proton number $Z$
and neutron number $N$. We plot the single neutron separation
energies of our calculations with TMA set (empty square) for Ne,
Na, Cl and Ar isotopes, together with the available experimental
values \cite{audi.95} (solid square) in Fig. 1. Good agreement
between experiment and our calculations can be clearly seen.
Calculations with TM2 set and NL3 set give practically the same
results. For Ne isotopes, our calculations give relative small
odd-even staggering than experiment except for $^{31}$Ne. The most
neutron-rich isotope ever observed for Ne isotopes is $^{34}$Ne
\cite{notani.02} and for Na isotopes it is $^{37}$Na
\cite{notani.02} , which are consistent with our present
calculations (see also Table. \ref{table1} and Table.
\ref{table2}). The single proton separation energies of our
calculations with TMA set (empty square) for Na and Ar isotopes,
together with the available experimental values \cite{audi.95}
(solid square) are plotted in Fig. 2. Once again, good agreement
between experiment and our calculations can be clearly seen. For
$^{21}$Na, the difference between experiment and our calculations
seems somewhat large. We should note that, from the definition of
$S_p$ (Eq. \ref{eq:sp}), the single proton separation energy for
$^{21}$Na is related to the binding energies of $^{21}$Na and
$^{20}$Ne. While in our calculations the calculated binding energy
for $^{20}$Ne is about 1.3 MeV smaller than experiment and for
$^{21}$Na is about 1.2 MeV larger than experiment. Thus a total
difference of 2.5 MeV is given to the single proton separation
energy for $^{21}$Na. For $^{35}$Ar, the relatively large
difference comes from the fact that the calculated binding energy
for $^{34}$Cl is about 2.5 MeV larger than experiment, while
calculated binding energy for $^{35}$Ar is about 1 MeV larger than
experiment. The last bound proton-rich isotope in Na isotopes is
predicted to be $^{20}$Na in our calculations, which has a single
proton separation energy 3.099 MeV. Its one neutron less neighbor,
$^{19}$Na, has almost a zero single proton separation energy
-0.013 MeV (see also Table. \ref{table1} and Table.\ref{table2}).
The last bound proton-rich isotope in Ar isotopes is predicted to
be $^{31}$Ar in our calculations, whose $S_p$ is 0.741 MeV (see
also Table. \ref{table3} and Table. \ref{table4}). All these
predictions are consistent with available experimental knowledge
\cite{woods.97}.

In conclusion, our calculations give a very good description of
single neutron and proton separation energies for Ne, Na, Cl and
Ar isotopes. All three calculations with different parameter sets,
TMA, TM2 and NL3, give essentially the same results. Therefore, we
can conclude that the binding energies and single neutron (proton)
separation energies are reproduced very well with TMA and NL3
parameter sets, which have been checked by many observables in the
entire mass region.

\subsection{Root mean square nuclear radius}
The root mean square neutron, proton, and matter radii are other
important basic physical quantities for nuclei in addition to the
single neutron (proton) separation energies. In the RMF theory,
the root mean square (rms) neutron, proton and matter radii can be
directly deduced from the neutron, proton and matter density
distributions, $\rho_n$, $\rho_p$ and $\rho_m$,
\begin{equation}
R_i=\langle r^2_i\rangle^{1/2}= \{\frac{\int \rho_i r^2d{\bf
r}}{\int \rho_i d{\bf r}}\}^{1/2},
\end{equation}
where the index $i\mbox{ }(=n, p, m)$ denotes the corresponding
neutron, proton and matter density distributions. First let us
have a look at the rms neutron and rms matter radii of Na and Ar
isotopes. In Fig. 3, we plot results of our calculations with TMA
set and the available experimental values for Na
\cite{suzuki.95,suzuki.98} and Ar isotopes \cite{ozawa.02}. For
neutron radii and matter radii of Na isotopes, we notice that our
calculations agree very well with experiment except for $^{22}$Na,
which has recently been attributed to an admixture of the isomeric
state in the beam \cite{suzuki.98}. For Ar isotopes, although the
mass dependence of the experimental data are reproduced quite
well, some discrepancies remain. First, the theoretical
predictions are at the upper limits of experimental values.
Calculations with TM2 set and NL3 set are not better either.
Second, the larger radii of $Z=N$ nuclei $^{36}$Ar than those of
its neighbors has been interpreted to be a possible alpha-cluster
structure \cite{ozawa.96}. Third, our calculations do not
reproduce the abrupt decrease of $^{37}$Ar and $^{38}$Ar. The
relatively larger radii of proton drip line nucleus $^{31}$Ar can
be understood easily if we note that $^{31}$Ar is the last bound
proton-rich isotope in Ar isotopes. Second we plot the charge
isotope shifts of our calculations with TMA set (empty square)
together with the experimental values (solid square) for Na
isotopes \cite{otten.89} and Ar isotopes \cite{klein.96} in Fig.
4. Calculations with TM2 set (solid circle) and results of HF+BCS
model \cite{gori.01} are also shown for comparison. For Na
isotopes, significant difference between the theoretical
predictions and the experimental values exist in two regions. The
first is $^{25}$Na, and the second is around the region of neutron
drip line $A\ge 29$. While for Ar isotopes, calculations with TM2
set agree very well with the experimental values within the error
bar except for $^{40}$Ar. For both isotopes, results with TM2 set
are closer to the experimental values. It is not surprising
because TM2 is a parameter set specially made for light nuclei.
For further comparison with other theories, we include the results
of the HF+BCS model \cite{gori.01} in Fig. 4. As for Ne isotopes,
the isotope shifts start to deviate from $^{26}$Ne to the lower
side except for $^{31}$Ne. On the other hand, the results for Ar
isotopes agree very well with experiment and are similar to those
of the RMF+BCS with TM2 parameter set except for $^{32}$Ar.

To conclude, the deformed RMF+BCS method describe the rms neutron
radii and matter radii very well. While for rms charge (proton)
radii, although the basic trend is reproduced quite well, the
results are not so satisfactory considering that we obtain the
parameter set by fitting the charge radii of certain nuclei.
However, we might have to pay more attention to the various
cluster phenomena, which make it difficult to apply the mean field
theory to light nuclei.

\subsection{Neutron skin and proton skin}
Different theories, both relativistic and non-relativistic ones,
have predicted the existence of proton skin and neutron skin since
long time ago, but only recently experimental physicists have
proved the theoretical predictions. In Fig. \ref{fig5.fig}, the
proton skin $\Delta R=R_p-R_n$ is plotted against the difference
between the proton and neutron separation energy $\Delta
S=S_p-S_n$. It can be clearly seen that $\Delta R$ has a strong
correlation with $\Delta S$. Thus, it is shown that the difference
between the proton and neutron Fermi energy is the driving force
for the creation of the skin phenomenon. Such a correlation has
already been predicted by RMF theory since a decade ago
\cite{tanihata.92}. For Na isotopes, calculations with NL3 set
predict smaller $\Delta S$ for $^{28-31}$Na, while calculations
with TMA and TM2 sets are in better agreement with experimental
$\Delta S$ values. For Ar isotopes, all three calculations failed
to reproduce the experimental observed proton-skin in $^{37}$Ar
and $^{38}$Ar. In Fig. \ref{fig6.fig}, we plot neutron skin
$R_n-R_p$ against mass number A for Ne, Na, Cl and Ar isotopes.
From Figure. \ref{fig6.fig}, it is quite clear that neutron
(proton) skin are quite common for neutron-rich (proton-rich)
nuclei.

In the Relativistic Mean Field theory, the formation of skin or
halo is usually explained by the picture of a core plus a few
occupied loosely bound (unbound) states of small angular momentum.
The halo in $^{11}$Li has been successfully reproduced in this
picture \cite{meng.96} where the occupations of 1p$_{1/2}$ and
2s$_{1/2}$ state are found to be important. Based on the similar
picture, giant-halos in $^{124-138}$Zr have been predicted by both
RCHB \cite {meng.98} method and resonant RMF-rBCS method
\cite{sand.03}.  In Ref. \cite{sand.03}, it has been shown that
3p$_{3/2}$, 2f$_{7/2}$ and 3p$_{1/2}$ states contribute most to
the formation of halo in $^{124}$Zr. In order to reproduce the
sudden increase of rms neutron (proton) radii, which is one
indication of halo structure, the Relativistic Mean Field model
must be solved in coordinate space. Although, compared with RCHB
method, quite similar two neutron separation energies are obtained
in the deformed RMF+BCS method \cite{geng.03}, the sudden increase
of rms neutron radii for $^{124-138}$Zr are not reproduced well.
Some improvements on the expansion method could improve the tail
behavior of resonant wave functions and solve this problem
finally. Possible candidates are the expansion method in the local
transformed Harmonic Oscillator basis \cite{stoitsov.98} and the
expansion method in a Woods-Saxon basis \cite{zhou.03}. This work
is under way now. However, as we can see from our calculations,
such a disadvantage of Harmonic Oscillator basis does not
influence our conclusions here.

\subsection{Deformation parameter}

Compared with binding energies and nuclear radii, predicted
deformation parameters of nuclei are quite different from model to
model. One reason is, of course, the possible shape coexistence.
Another reason is that deformation parameters are more sensitive
to model details. On the experimental side, we can obtain the
information of nuclear deformation from measurements of
$B(E2)\uparrow$ values. The $B(E2)\uparrow$ values are basic
experimental quantities that do not depend on nuclear models.
Assuming a uniform charge distribution out to the distance
$R(\theta,\phi)$ and zero charge beyond, $\beta_{2p}$
\cite{raman.01,geng.03} is related to $B(E2)\uparrow$ by
 \begin{equation}
 \beta_{2p}=(4\pi/3Z R^2_0)[B(E2)\uparrow/e^2]^{1/2},
 \end{equation}
where $R_0$ has been taken to be $1.2A^{1/3}$ fm and
$B(E2)\uparrow$ is in units of $e^2b^2$. $B(E2)\uparrow$ values
have been measured mainly for some even-even nuclei.

In Fig. \ref{fig7.fig}, we plot the deformation parameter
$\beta_{2p(m)}$ for Ne, Na, Cl and Ar isotopes against mass number
A. For our RMF+BCS calculations, we plot $\beta_{2p}$ because as
we can see from Eq. (13) that what we obtain from $B(E2)\uparrow$
is actually $\beta_{2p}$, not $\beta_{2m}$. For the Finite Range
Droplet Model (FRDM) \cite{moller.95}, the plotted deformation
parameter is $\beta_{2m}$. Except for exotic nuclei with extreme
N/Z ratios, we note that these two values are very close to each
other (see also Table. \ref{table1} $\sim$ \ref{table4}). From
Fig. \ref{fig7.fig}, we see that experimental values show that Ne
isotopes are super deformed ($\beta_{2p}\approx0.6$) in the region
$A=18\sim28$. Except for $^{20}$Ne and $^{22}$Ne, our calculations
failed to reproduce $^{18}$Ne, $^{24}$Ne, $^{26}$Ne and $^{28}$Ne.
For Na isotopes, both our calculations and FRDM predict prolate
shapes. For Cl isotopes, a transition between prolate shapes and
oblate shapes have been predicted by both methods. For Ar
isotopes, our calculations predict oblate shapes for most nuclei
while FRDM predict spherical shapes. Except for $^{38}$Ar, our
predictions agree very well with experimental values. We should
mention here that, because from the $B(E2)\uparrow$ values, we can
not distinguish prolate or oblate shapes, we just put a random
sign before the experimental value to make it close to our
predictions. This does not change the essence of our comparison.

\section{Summary}

We have studied light nuclei, in particular, Ne, Na, Cl and Ar
isotopes, in the framework of the deformed RMF+BCS method. The
deformation is treated by using the expansion method in the
deformed Harmonic Oscillator basis.  We have used TMA, TM2 and NL3
effective interactions in the RMF Lagrangian. In addition we have
treated the pairing correlations in terms of a density-independent
delta-function interaction.

We have calculated neutron (proton) separation energies,
quadrupole deformations, nuclear neutron (proton) and matter
radii.  All the results are presented in the form of tables
(Table. I $\sim$ IV).  We have compared the calculated results
with three parameter sets, TMA, TM2 and NL3, with the available
experimental values. Both TMA and NL3 parameter sets are often
used in the RMF theory in the entire mass region from the proton
drip line to the neutron drip line. We have used TM2 parameter set
also, which is extracted particularly for light nuclei. All the
three parameter sets provide similar results in our present
calculations.

The single neutron (proton) separation energies are reproduced
very well within the deformed RMF+BCS method with all the three
parameter sets. The general trends of the rms neutron radii come
out to be also very good. However, the rms proton radii seem to be
generally overestimated in the deformed RMF+BCS method in
comparison with experimental values. We have extracted also the
nuclear deformations by the minimization method. In general, the
calculated results are in good agreement with experimental values
extracted from the $BE(2)\uparrow$ values. The discrepancy on the
proton radii for these light nuclei may be related with the
softness of the light nuclei in deformation and in
alpha-clustering, which are not included in the mean field
Lagrangian. We have to study the effect of the softness in future
works

\section{Acknowledgments}

L.S. Geng is grateful to the Monkasho fellowship for supporting
his stay at Research Center for Nuclear Physics where this work is
done .

\newpage

\begin{figure}[t]
\centering
\includegraphics[scale=0.6]{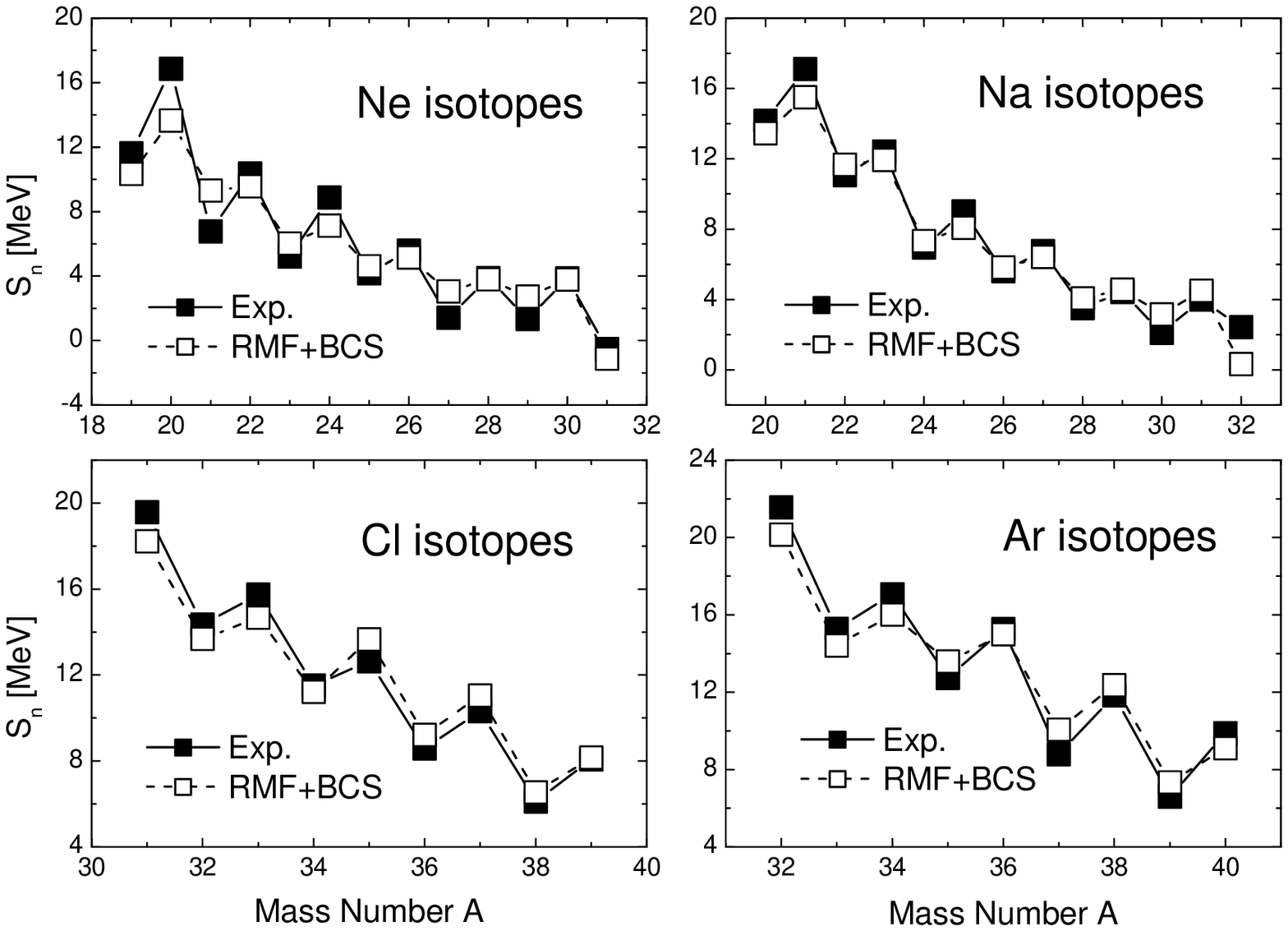}
\caption{\label{fig1.fig} The single neutron separation energis,
S$_n$, of Ne, Na, Cl and Ar isotopes. Results obtained from the
deformed RMF+BCS calculations with TMA set are compared with
available experimental data \cite{audi.95}.}
\end{figure}

\newpage

\begin{figure}[t]
\centering
\includegraphics[scale=0.6]{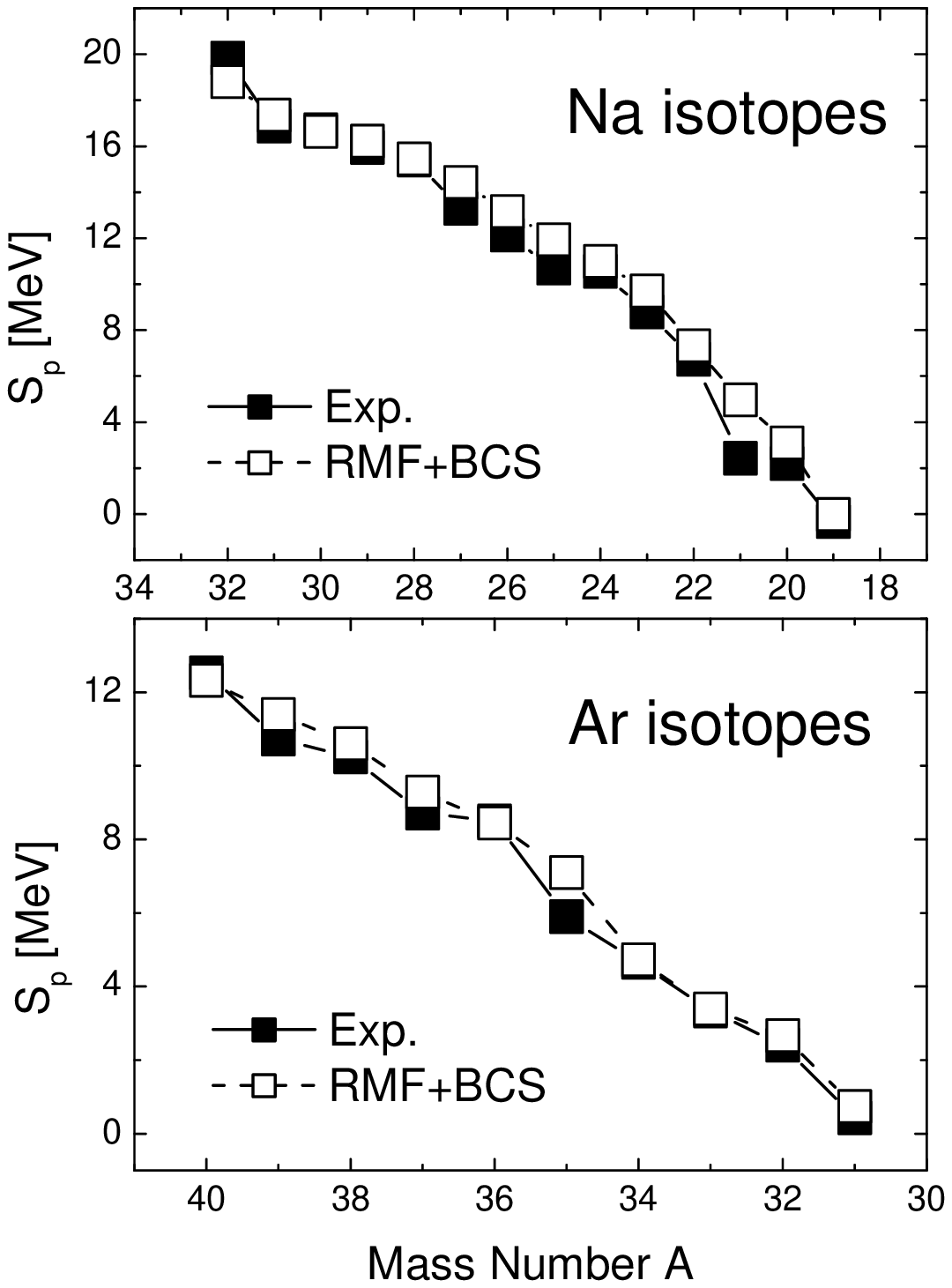}
\caption{\label{fig2.fig} The single proton separation energies,
S$_p$, of Na and Ar isotopes. Results obtained from the deformed
RMF+BCS calculations with TMA set are compared with available
experimental data \cite{audi.95}.}
\end{figure}

\newpage

\begin{figure}[t]
\centering
\includegraphics[scale=0.55]{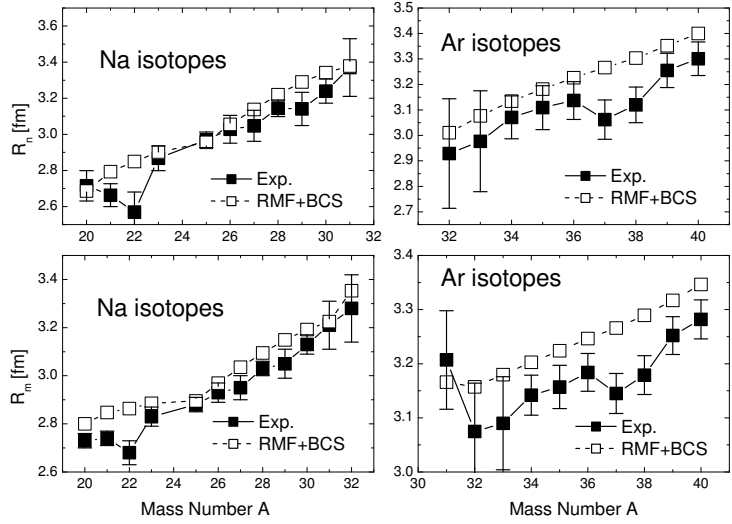}
\caption{\label{fig3.fig} The rms neutron radii, $R_n$, and rms
matter radii, $R_m$, of Na and Ar isotopes. Results obtained from
the deformed RMF+BCS calculations with TMA set are compared with
available experimental data \cite{suzuki.95,suzuki.98,ozawa.02}.}
\end{figure}

\newpage

\begin{figure}[t]
\centering
\includegraphics[scale=0.55]{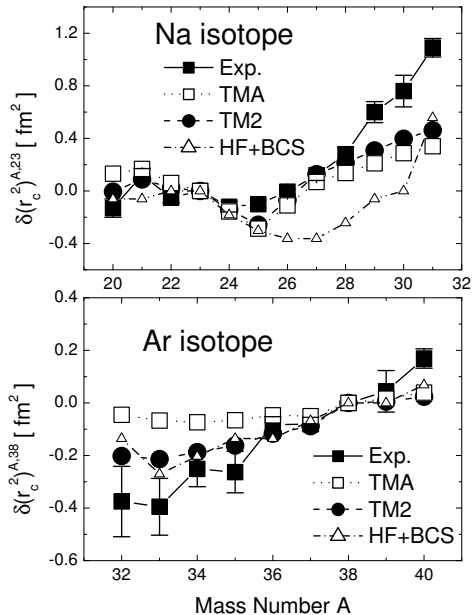}
\caption{\label{fig4.fig} The charge isotope shifts of Na and Ar
isotopes. Results obtained from the deformed RMF+BCS calculations
with TMA set and TM2 set are compared with available experimental
data \cite{otten.89,klein.96} and results of HF+BCS model
\cite{gori.01}.}
\end{figure}

\newpage

\begin{figure}[t] \centering
\begin{minipage}[c]{0.5\linewidth}
\includegraphics[scale=0.35]{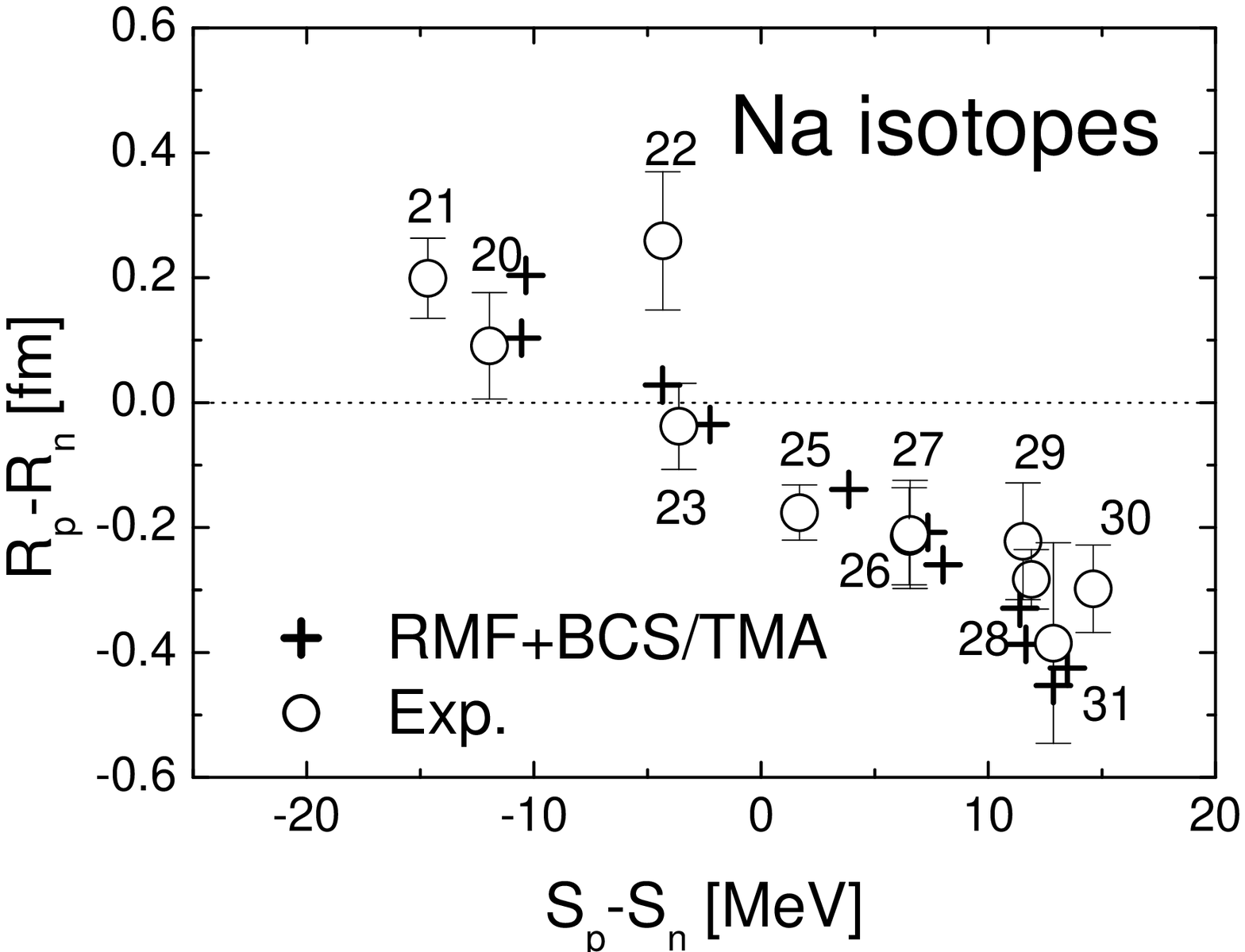}
\end{minipage}%
\begin{minipage}[c]{0.5\linewidth}
\includegraphics[scale=0.35]{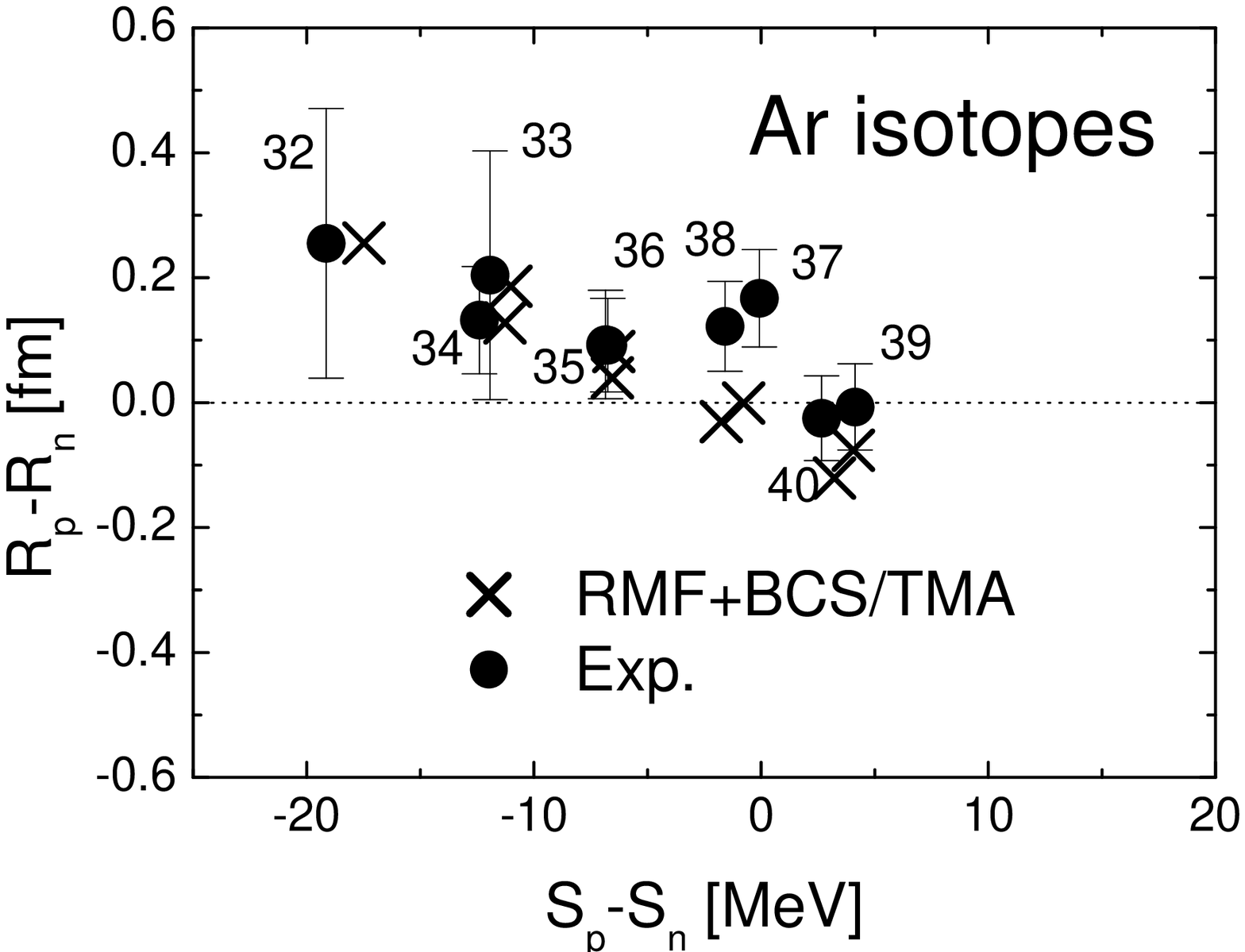}
\end{minipage}
\begin{minipage}[c]{0.5\linewidth}
\includegraphics[scale=0.35]{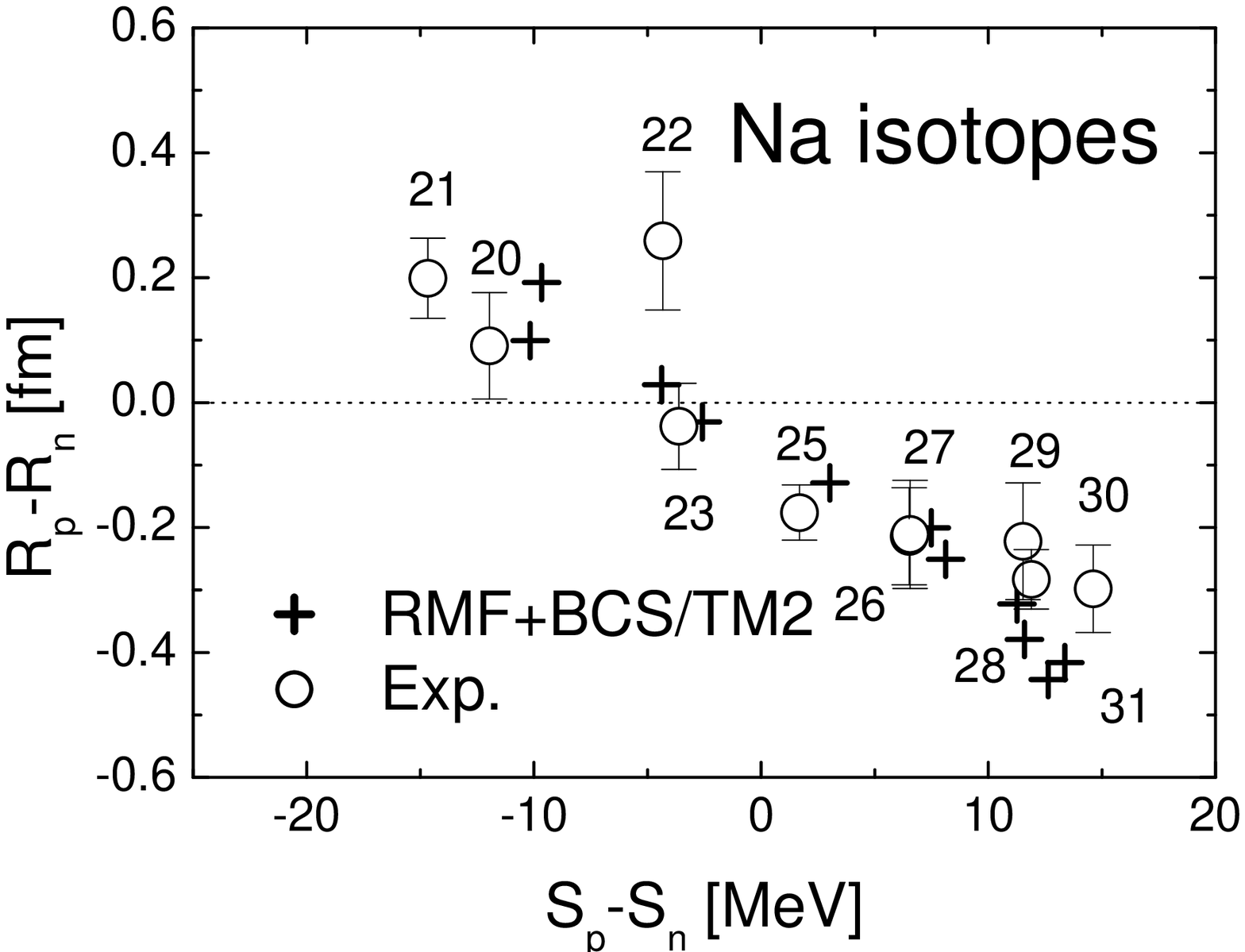}
\end{minipage}%
\begin{minipage}[c]{0.5\linewidth}
\includegraphics[scale=0.35]{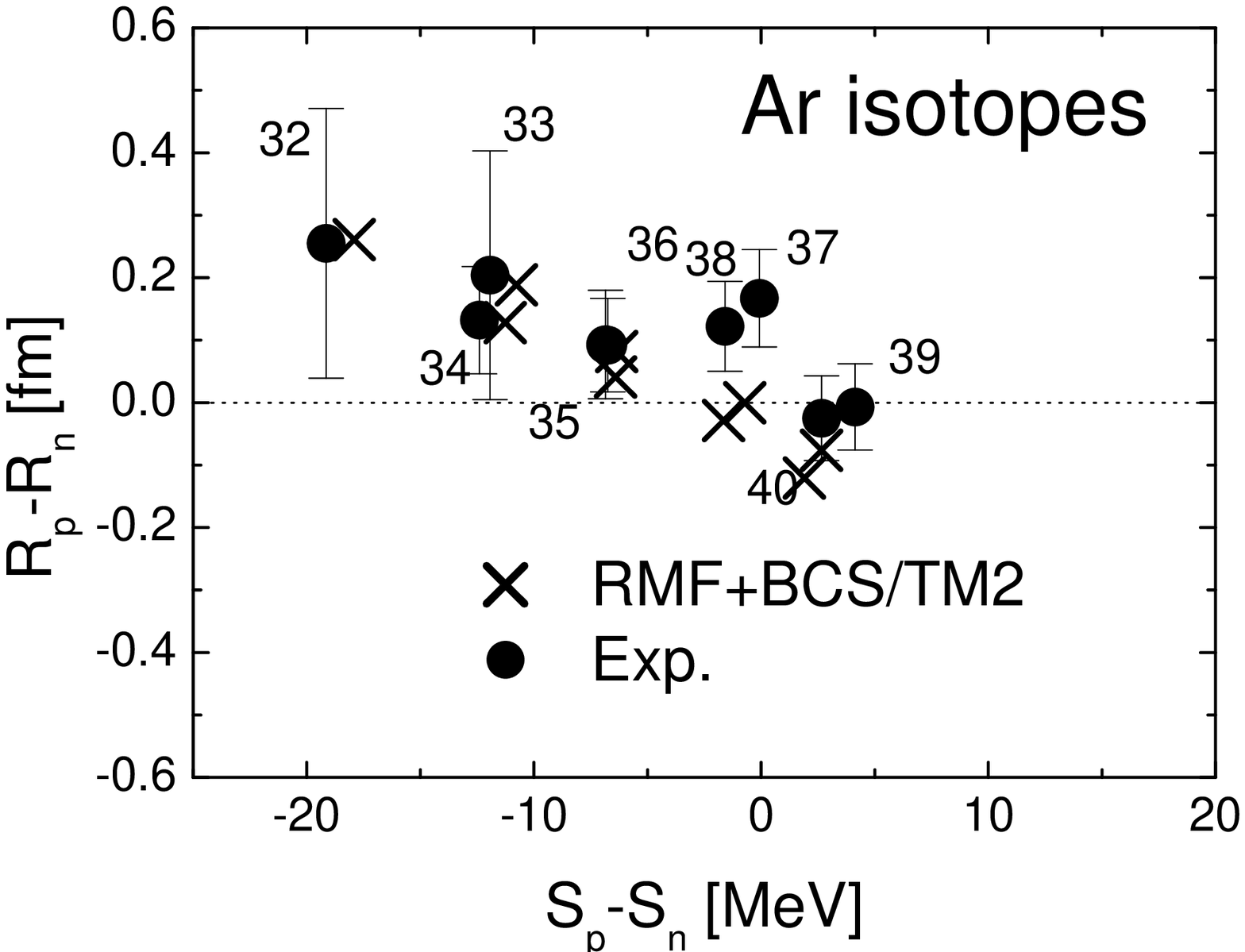}
\end{minipage}
\begin{minipage}[c]{0.5\linewidth}
\includegraphics[scale=0.35]{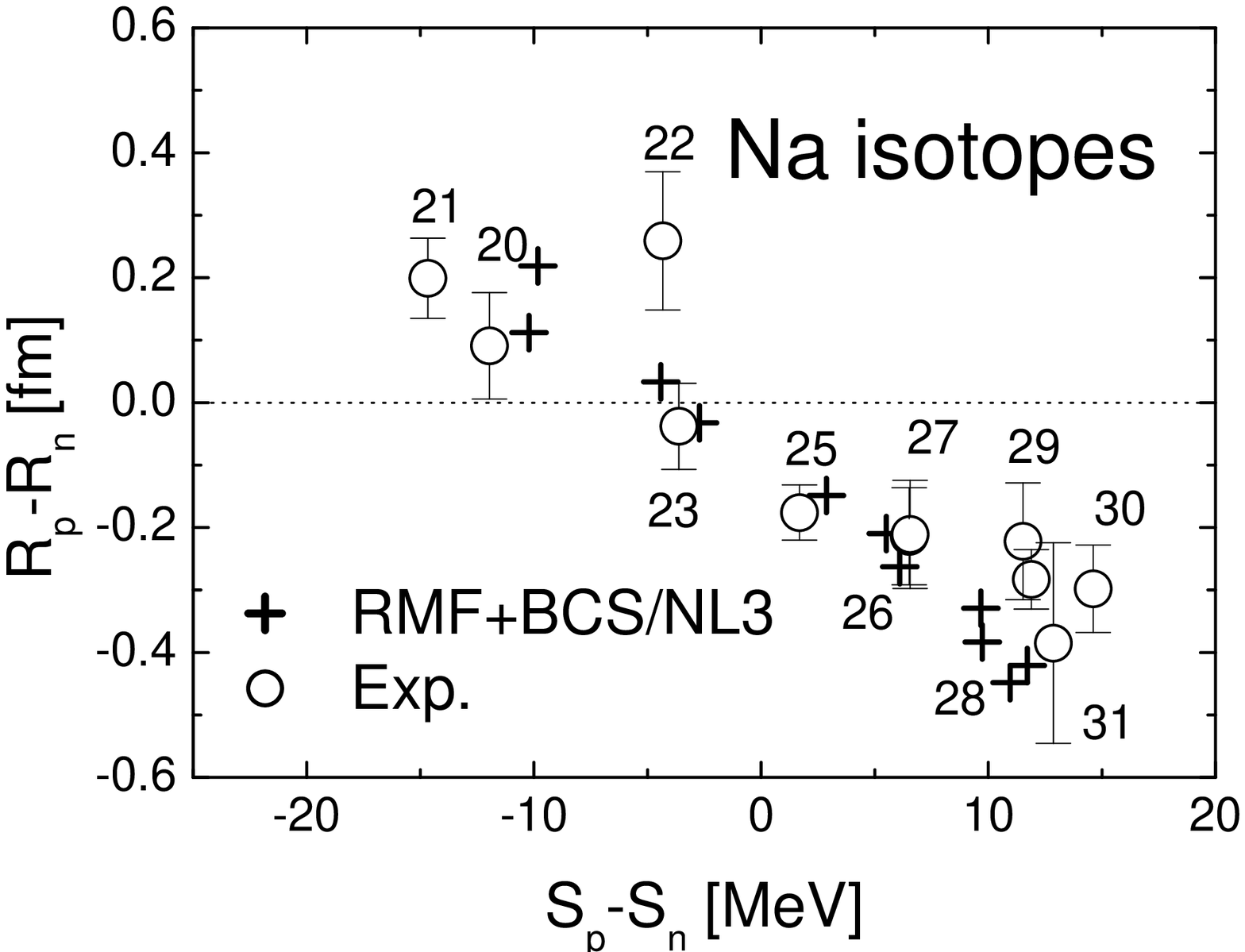}
\end{minipage}%
\begin{minipage}[c]{0.5\linewidth}
\includegraphics[scale=0.35]{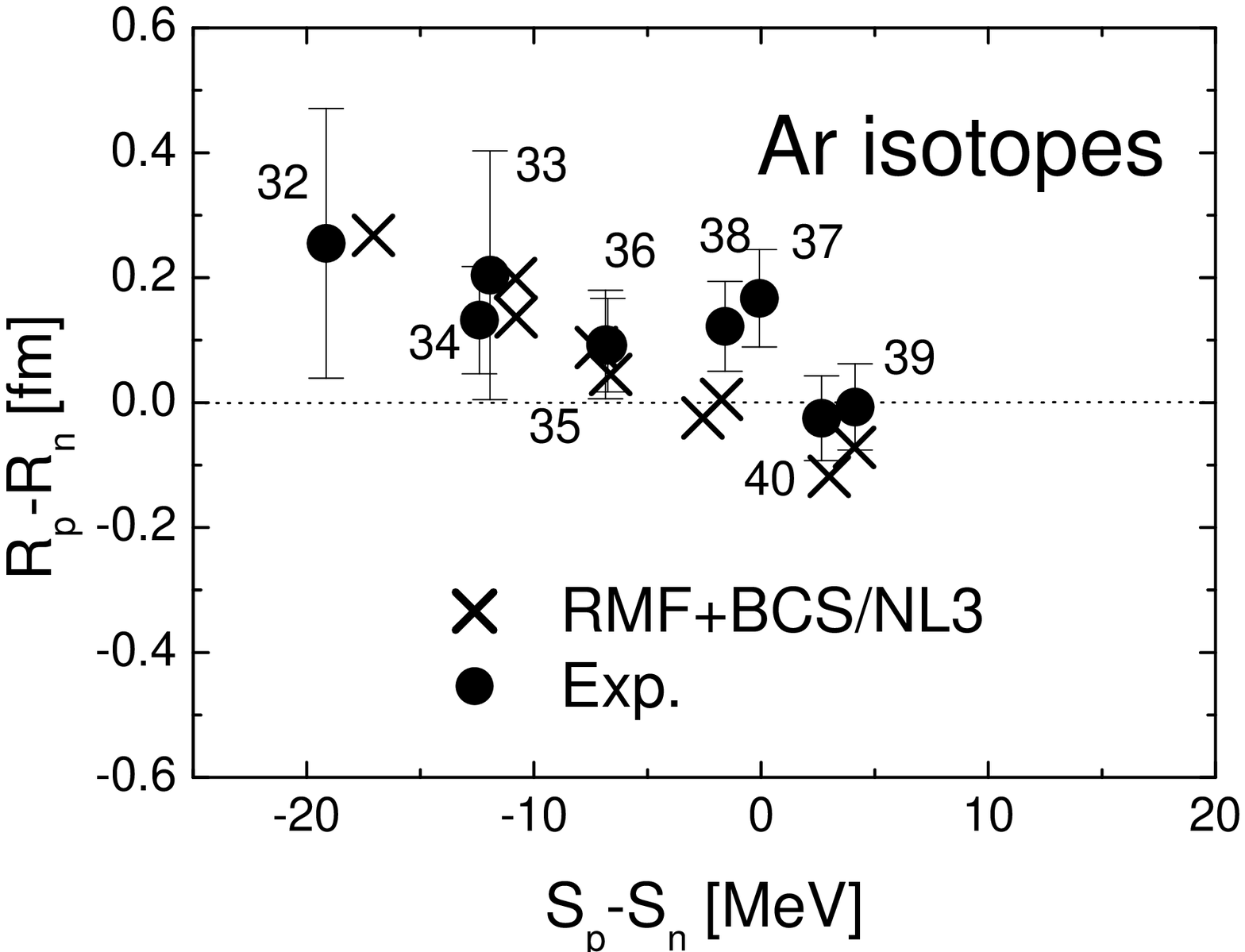}
\end{minipage}
\caption{\label{fig5.fig}The proton skins, $R_p-R_n$, plotted as
functions of the difference between the proton and neutron
separation energy, $S_p-S_n$. We show the theoretical results with
all the three parameter sets, TMA, TM2 and NL3, from top to
bottom. The corresponding experimental data are taken from Ref.
\cite{suzuki.95,suzuki.98,ozawa.02,audi.95}.}
\end{figure}

\newpage

\begin{figure}[t]
\centering
\includegraphics[scale=0.6]{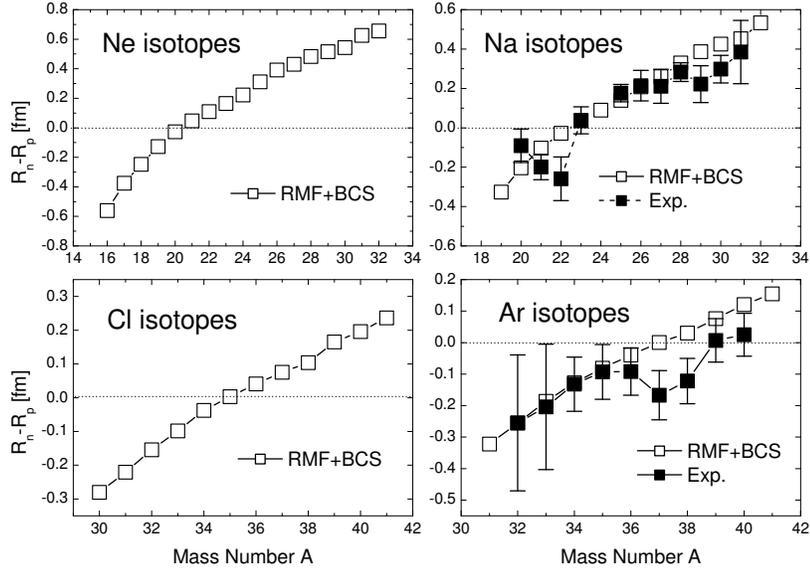}
\caption{\label{fig6.fig}The neutron skins, $R_n-R_p$, of Ne, Na,
Cl and Ar isotopes. Results obtained from the deformed RMF+BCS
calculations with TMA set are compared with available experimental
data \cite{suzuki.95,suzuki.98,ozawa.02}. }
\end{figure}

\newpage

\begin{figure}[t]
\centering
\includegraphics[scale=0.6]{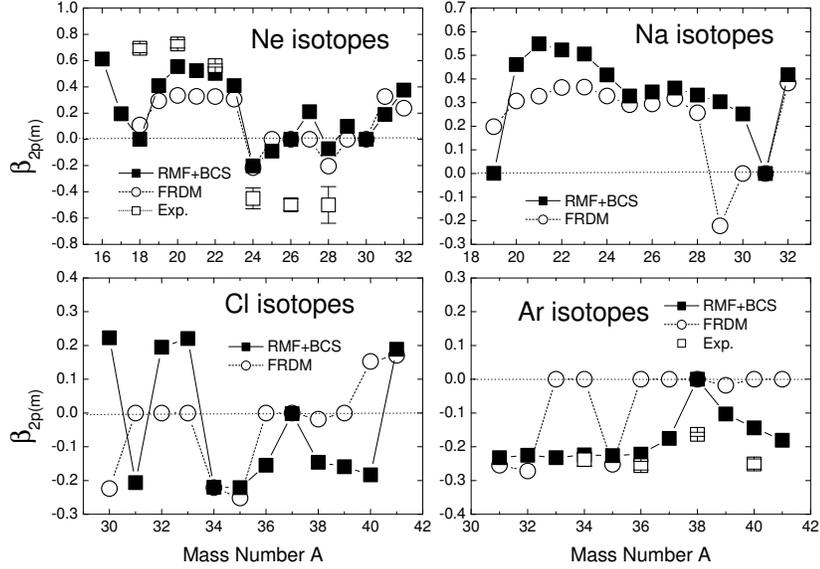}
\caption{\label{fig7.fig}The deformation parameters,
$\beta_{2p(m)}$, of Ne, Na, Cl and Ar isotopes. Results obtained
from the deformed RMF+BCS calculations with TMA set are compared
with available experimental data \cite{raman.01} and results of
FDRM model \cite{moller.95}. }
\end{figure}

 \setlength{\tabcolsep}{0.25
 em}
\begin{table}[htbp]
\caption{The ground state properties of Ne isotopes ($Z=10$)
calculated with the parameter set TMA. Listed are the total
binding energy, $B$, the binding energy per nucleon, $B/A$ ,
charge, neutron, proton, and matter root mean square radii, $R_c$,
$R_n$, $R_p$ and $R_m$, and the quadrupole deformation parameter
for the neutron, proton and matter distributions, $\beta_{2n}$,
$\beta_{2p}$ and $\beta_{2m}$, with $A$ the mass number and $N$
the neutron number.}
\begin{center}\label{table1}
\begin{tabular}{cc@{\hspace{2ex}}|@{\hspace{2ex}}cc@{\hspace{2ex}}|@{\hspace{2ex}}cccc@{\hspace{2ex}}|@{\hspace{2ex}}ccc}
\hline\hline
$A$&$N$&$B$&$B/A$&$R_c$&$R_n$&$R_p$&$R_m$&$\beta_{2n}$&$\beta_{2p}$&$\beta_{2m}$\\
 \hline\hline
 16&6&99.115&6.196&3.0845&2.4177&2.9789&2.7818&0.249&0.612&0.476\\
 17&7&116.263&6.839&2.9811&2.4967&2.8717&2.7236&0.076&0.195&0.146\\
 18&8&135.379&7.521&2.9196&2.5598&2.8078&2.7004&0.000&0.000&0.000\\
  19&9&145.724&7.670&2.9194&2.6811&2.8077&2.7485&0.308&0.409&0.361\\
  20&10&159.367&7.968&2.9392&2.8011&2.8282&2.8147&0.541&0.554&0.548\\
  21&11&168.666&8.032&2.9251&2.8604&2.8135&2.8382&0.531&0.523&0.528\\
  22&12&178.234&8.102&2.9176&2.9155&2.8057&2.8661&0.527&0.502&0.515\\
  23&13&184.257&8.011&2.8942&2.9462&2.7814&2.8757&0.375&0.409&0.390\\
  24&14&191.392&7.975&2.8637&2.9726&2.7497&2.8818&-0.235&-0.203&-0.222\\
  25&15&195.991&7.840&2.8672&3.0642&2.7533&2.9438&-0.086&-0.090&-0.088\\
  26&16&201.148&7.736&2.8825&3.1613&2.7693&3.0165&0.000&0.000&0.000\\
  27&17&204.194&7.563&2.9250&3.2443&2.8135&3.0917&0.181&0.210&0.191\\
  28&18&207.988&7.428&2.9392&3.3118&2.8283&3.1476&-0.069&-0.071&-0.070\\
  29&19&210.725&7.266&2.9643&3.3695&2.8543&3.2012&0.071&0.099&0.081\\
  30&20&214.508&7.150&2.9863&3.4206&2.8772&3.2496&0.000&0.000&0.000\\
  31&21&213.381&6.883&3.0102&3.5274&2.9020&3.3385&0.210&0.190&0.203\\
  32&22&215.128&6.723&3.0655&3.6155&2.9593&3.4240&0.388&0.376&0.384\\
  33&23&215.486&6.530&3.0905&3.6820&2.9851&3.4856&0.453&0.407&0.439\\
  34&24&216.331&6.363&3.1159&3.7446&3.0115&3.5447&0.513&0.436&0.491\\
 \hline\hline
\end{tabular}
\end{center}
\end{table}

\begin{table}[htbp]
\caption{The same as Table \ref{table1}, but for Na isotopes
($Z=11$).}
\begin{center}\label{table2}
\begin{tabular}{cc@{\hspace{2ex}}|@{\hspace{2ex}}cc@{\hspace{2ex}}|@{\hspace{2ex}}cccc@{\hspace{2ex}}|@{\hspace{2ex}}ccc}
\hline\hline
$A$&$N$&$B$&$B/A$&$R_c$&$R_n$&$R_p$&$R_m$&$\beta_{2n}$&$\beta_{2p}$&$\beta_{2m}$\\
 \hline\hline
 18&7&115.054&6.392&3.0876&2.5260&2.9822&2.8136&0.193&0.503&0.383\\
 19&8&135.366&7.125&2.9995&2.5642&2.8908&2.7580&0.000&0.001&0.001\\
 20&9&148.823&7.441&2.9985&2.6863&2.8899&2.8001&0.325&0.461&0.400\\
 21&10&164.329&7.825&3.0044&2.7928&2.8960&2.8473&0.516&0.549&0.533\\
 22&11&175.991&8.000&2.9865&2.8493&2.8774&2.8634&0.512&0.523&0.517\\
 23&12&187.934&8.171&2.9764&2.9020&2.8668&2.8853&0.511&0.506&0.508\\
 24&13&195.248&8.135&2.9501&2.9297&2.8395&2.8887&0.369&0.417&0.391\\
 25&14&203.333&8.133&2.9279&2.9557&2.8165&2.8953&0.243&0.328&0.280\\
 26&15&209.157&8.044&2.9579&3.0555&2.8477&2.9694&0.288&0.345&0.312\\
 27&16&215.574&7.984&2.9878&3.1383&2.8787&3.0352&0.333&0.362&0.345\\
 28&17&219.644&7.844&2.9992&3.2198&2.8905&3.0946&0.259&0.332&0.288\\
 29&18&224.209&7.731&3.0116&3.2903&2.9034&3.1491&0.198&0.304&0.238\\
 30&19&227.366&7.579&3.0243&3.3416&2.9165&3.1923&0.126&0.252&0.172\\
 31&20&231.868&7.480&3.0330&3.3784&2.9256&3.2250&0.000&0.000&0.000\\
 32&21&232.198&7.256&3.0989&3.5270&2.9938&3.3533&0.433&0.418&0.428\\
 33&22&235.422&7.134&3.1079&3.5611&3.0031&3.3853&0.361&0.371&0.364\\
 34&23&236.869&6.967&3.1330&3.6225&3.0292&3.4418&0.419&0.404&0.414\\
 35&24&238.780&6.822&3.1588&3.6815&3.0558&3.4969&0.476&0.435&0.463\\
 36&25&238.540&6.626&3.1726&3.7808&3.0701&3.5786&0.531&0.434&0.501\\
 37&26&238.831&6.455&3.1840&3.8488&3.0818&3.6377&0.540&0.426&0.506\\
 \hline\hline
\end{tabular}
\end{center}
\end{table}

\begin{table}[htbp]
\caption{The same as Table \ref{table1}, but for Cl isotopes
($Z=17$).}
\begin{center}\label{table3}
\begin{tabular}{cc@{\hspace{2ex}}|@{\hspace{2ex}}cc@{\hspace{2ex}}|@{\hspace{2ex}}cccc@{\hspace{2ex}}|@{\hspace{2ex}}ccc}
\hline\hline
$A$&$N$&$B$&$B/A$&$R_c$&$R_n$&$R_p$&$R_m$&$\beta_{2n}$&$\beta_{2p}$&$\beta_{2m}$\\
 \hline\hline
28&11&187.833&6.708&3.4060&2.8804&3.3107&3.1487&0.336&0.271&0.297\\
29&12&208.281&7.182&3.3718&2.9316&3.2755&3.1378&0.352&0.287&0.314\\
30&13&225.271&7.509&3.3264&2.9484&3.2287&3.1104&0.233&0.223&0.227\\
31&14&243.494&7.855&3.3031&2.9843&3.2048&3.1072&-0.193&-0.206&-0.200\\
32&15&257.174&8.037&3.2968&3.0439&3.1983&3.1269&0.198&0.195&0.196\\
33&16&271.847&8.238&3.3053&3.1084&3.2070&3.1596&0.250&0.221&0.235\\
34&17&283.061&8.325&3.3055&3.1699&3.2073&3.1886&-0.204&-0.220&-0.212\\
35&18&296.711&8.477&3.3140&3.2188&3.2160&3.2174&-0.218&-0.221&-0.220\\
36&19&305.930&8.498&3.3145&3.2574&3.2165&3.2382&-0.117&-0.155&-0.135\\
37&20&316.981&8.567&3.3211&3.2986&3.2233&3.2642&-0.001&-0.002&-0.002\\
38&21&323.520&8.514&3.3255&3.3515&3.2278&3.2968&-0.102&-0.146&-0.122\\
39&22&331.685&8.505&3.3328&3.4005&3.2354&3.3295&-0.126&-0.159&-0.140\\
40&23&337.720&8.443&3.3433&3.4420&3.2462&3.3602&-0.164&-0.183&-0.172\\
41&24&345.731&8.432&3.3498&3.4892&3.2529&3.3932&0.219&0.189&0.207\\
 \hline\hline
\end{tabular}
\end{center}
\end{table}

\begin{table}[htbp]
\caption{The same as Table \ref{table1}, but for Ar isotopes
($Z=18$).}
\begin{center}\label{table4}
\begin{tabular}{cc@{\hspace{2ex}}|@{\hspace{2ex}}cc@{\hspace{2ex}}|@{\hspace{2ex}}cccc@{\hspace{2ex}}|@{\hspace{2ex}}ccc}
\hline\hline
$A$&$N$&$B$&$B/A$&$R_c$&$R_n$&$R_p$&$R_m$&$\beta_{2n}$&$\beta_{2p}$&$\beta_{2m}$\\
 \hline\hline
29&11&186.450&6.429&3.4869&2.8961&3.3939&3.2142&0.317&0.212&0.252\\
30&12&207.631&6.921&3.4499&2.9427&3.3559&3.1970&0.325&0.217&0.260\\
31&13&226.012&7.291&3.3926&2.9750&3.2970&3.1659&-0.208&-0.232&-0.222\\
32&14&246.157&7.692&3.3631&3.0111&3.2665&3.1573&-0.203&-0.225&-0.215\\
33&15&260.569&7.896&3.3598&3.0769&3.2631&3.1799&-0.214&-0.232&-0.224\\
34&16&276.577&8.135&3.3589&3.1345&3.2622&3.2027&-0.207&-0.224&-0.216\\
35&17&290.163&8.290&3.3600&3.1818&3.2634&3.2240&-0.216&-0.226&-0.221\\
36&18&305.154&8.477&3.3628&3.2264&3.2662&3.2464&-0.215&-0.222&-0.218\\
37&19&315.220&8.519&3.3623&3.2663&3.2657&3.26660&-0.217&-0.175&-0.151\\
38&20&327.581&8.621&3.3698&3.3038&3.2735&3.2895&0.000&0.000&0.000\\
39&21&334.913&8.588&3.3722&3.3521&3.2759&3.3171&-0.085&-0.103&-0.093\\
40&22&344.011&8.600&3.3756&3.4003&3.2795&3.3464&-0.115&-0.144&-0.128\\
41&23&351.032&8.562&3.3830&3.4418&3.2870&3.3747&-0.159&-0.181&-0.168\\
 \hline\hline
\end{tabular}
\end{center}
\end{table}

\end{document}